# Reconfigurable Integrated Photonic Chips as Dual-Purpose Neuromorphic Accelerators and Physical Unclonable Functions

**GEORGE SARANTOGLOU,[1] FRANCESCO DA ROS,[2] KOSTAS SOZOS[3], ADONIS BOGRIS[3], CHARIS MESARITAKIS[1*]**

[1]*Department of Biomedical Engineering, University of West Attica, Agiou Spiridonos,12243, Egaleo, Greece.*
[2] *Department of Electrical and Photonics Engineering, Technical University of Denmark, Ørsetds Plads Building 343, DK-2800 Kgs. Lyngby, Denmark.*
[3] *Department of Informatics & Computer Engineering, University of West Attica, Agiou Spiridonos 12243, Egaleo, Athens, Greece*
*\*cmesar@uniwa.gr*

**In this work, we experimentally validate the dual use of a reconfigurable photonic integrated mesh as a neuromorphic accelerator, targeting signal equalization, and as a physical unclonable function offering authentication at the hardware level. The processing node is an optical spectrum slicing self-coherent transceiver targeting the mitigation of dispersion impairments of an intensity detected QPSK signal, after 25 km of transmission at 32 Gbaud. Unavoidable fabrication related imperfections of the nodes, such as waveguide roughness, can act as "fingerprints" of the device, and, during neuromorphic processing, result in unique weights at the digital back-end during signal equalization. Extracted security metrics offer low false positive/negative probability <10$^{-8}$ for the generated responses, confirming un-clonability, whereas bit-error-ratio for the QPSK equalization task was always below the hardware forward error correction limit. The experimental results substantiate the capability of the proposed scheme to simultaneously act as an accelerator and as a security token.**

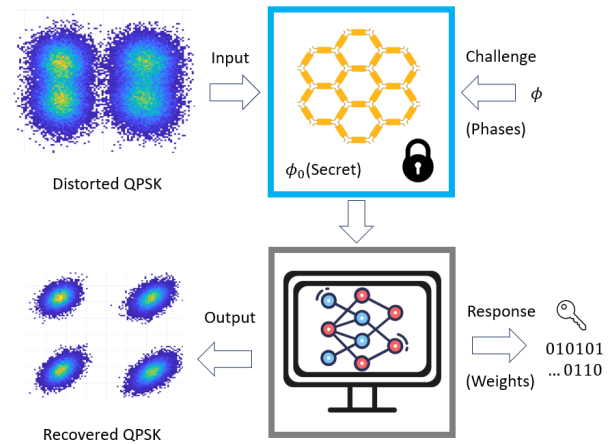

Fig. 1: The concept of an nPUF. On the left the input and the output of the hybrid (photonic and digital) neural network is presented. On the right, the challenge in the form of the phase values and the response (binarized weights) are presented.

Neuromorphic Computing is a promising alternative to conventional Von Neumann architectures in the ecosystem of machine learning [1]. One particularly promising pillar in this landscape are photonic hardware accelerators (PHAs), where processing is achieved by optical inference through reprogrammable structures, inspired by the architecture of neural networks [2]. The combination of wavelength division multiplexing and high bandwidth renders their use appealing in multiple tasks such signal equalization [3] or as convolutional kernels [4].

At the same time, the field of edge computing has started demanding the design of novel security mechanisms for device authentication and/or data encryption in the framework of trusted information and communication technology (ICT) systems. The foundation of most security schemes resides on the integrity and confidentiality of a secret key. These keys on the other hand are stored in non-volatile memory elements that are vulnerable to hardware/tampering attacks [5]. A cost-effective alternative to key generation and storage are physical unclonable functions (PUFs), where the unique and unpredictable physical properties of a device are exploited for key generation (response) after interrogation with a pre-determined input (challenge). Electronic PUFs such as ring oscillators, SRAM and memristors have decade long history [6]. Optical, scattering-based PUFs are considered more secure, but require bulky components [7]. Integrated photonic PUFs aim to inherit the high dimensionality of the scatter-based PUFs without the miniaturization limitations, but they require additional electro-optic circuitry to probe and interrogate them [8,9]. In our previous works [10] [11], we have proposed a neuromorphic-based PUF (nPUF), where a photonic chip could act as an integrated PUF and at the same time as a neuromorphic processor. Therefore, the key advantage in this scenario is that PUF responses are generated intrinsically as part of the acceleration process. In this work, this concept is validated experimentally, for the first time, using a commercial reconfigurable photonic integrated silicon chip (SmartLight) [12]. The chip was configured to realize a recurrent optical spectrum slicing (ROSS) receiver, similar to [3], whereas the target task was dispersion compensation for a self-coherent 32 Gbaud QPSK receiver after 25 km of transmission in the C-band [3]. Results demonstrate that the same structure can offer HD-FEC compatible performance as PHA and $10^{-8}$ probability of cloning as PUF.

The proposed concept is presented in Fig.1 and can be summarized as follows, the incoming signal- distorted QPSK- is split and fed to ROSS filters (nodes), realized through a hexagonal mesh of Mach Zehnder Interferometers (MZI) [12]. The nodes offer a two-fold operation; on one hand they convert phase information to the amplitude domain, allowing direct detection [13]. The second operation is that by slicing the signal at different frequencies, independent processing of each section is enabled, minimizing dispersion induced power fading [14]. The photocurrent from each ROSS node is sampled and the digital data are sent to a typical feed forward equalizer (FFE), where upon training, the regression coefficients are derived (the digital weights). Interestingly, these digital weights, as shown in [10] are irrevocably governed by the physical properties of the integrated mesh. In our case, these properties correspond to unintentional fabrication imperfections of the waveguides, due to e.g. sidewall roughness. These imperfections manifest as refractive index variations that follow a uniform distribution [15] leading, in turn, to passive phase offsets $\phi_0$ at each MZI, in the range $(0, 2\pi)$ [12]. This means that by using different MZIs (in the same chip) or using different chips so as to realize the ROSS nodes, the target task can be addressed (low bit error ratio – BER) but in the digital back-end, a different set of weights would be generated in each case. Therefore, in the proposed scheme, the intrinsic phase randomness ($\Phi_0$) can be considered as the PUF's physical secret, whereas the digital weights as the PUF's unique responses. Following this principle of operation, the digital weights project $\Phi_0$ can be used as an authentication token for device attestation, like in weak PUF scenarios [10]. Alternatively, during PIC reconfiguration, the user can apply specific phase shifts ($\Phi$) at the thermo-optic shifter of each MZI in the mesh, aiming to regulate their transmittance. The effective phase ($\phi_{eff}$), in this case, constitutes of both $\Phi$ and $\Phi_0$, $\phi_{eff} = \phi + \phi_0$.

Obviously, different $\Phi$ settings can evoke a non-linear (due to cosine relationship of the modulator) and unpredictable (due to $\Phi_0$) change in the corresponding weights. In this context, user driven $\Phi$ can act as a PUF challenge to the system, following a strong PUF scenario. Similarly, if the same structure is utilized, but a different task is assumed - different propagation parameters (distance, dispersion etc.), different modulation format/rate - the digital weights will also alter. Thus, these parameters, although not in this work, can act as part of the challenge for our scheme. In the proposed concept, applied $\Phi$ regulates each node's frequency detuning with respect to the carrier's frequency. This is a key hyperparameter for the signal equalization task, as it is shown in [3]. Here, by choosing a different set of $\Phi$ (challenge) so as to tune the system for minimum BER, at the same time we evoke specific digital weights (responses) offering a true dual modality intrinsic PUF, that does not need additional circuitry for response interrogation. Obviously, due to the different underlying phase biases ($\Phi_0$) at each MZI, the application of the same challenge ($\Phi$) to different devices/units in the device, would result in a different $\Phi_{eff}$ and in turn, to different responses and equalization performance [11].

In order to validate the aforementioned concept the experimental setup presented in Fig.2a was realized. A laser (λ=1550 nm) is partially modulated through an I/Q modulator with 55000 symbols using QPSK at 32 Gbaud to provide a signal with a residual carrier [3]. The signal propagates through a 25 km long optical fiber, which introduces chromatic dispersion. Its output is driven to an erbium-doped-fiber-amplifier (EDFA) that raises the optical input power before the SmartLight processor. SmartLight is programmed as a Mach Zehnder Delayed Interferometer (MZDI) in a loop, according to the ROSS paradigm (see Fig.2a). The bandwidth of the MZDI is 11.1 GHz, whereas the loop offers a time-delay of 22.2 ps. The bandwidth of the ROSS node is lower compared to the input signal (32 GHz), allowing for spectral slicing [14]. The insertion losses of the SmartLight structure are 22.5 dB, necessitating the use of a second EDFA followed by a 1-nm wide optical filter to remove noise. Afterwards, 1% of optical power is driven to an optical spectrum analyzer (OSA) and 99% to a 50 GHz photodiode. A digital storage oscilloscope (DSO) tracks the signal at 80 GSa/s. The signal is resampled down to 1 sample per symbol. Since the experiment is SNR-limited due to the 22 dB minimum losses of the Smartlight, averaging is applied in a progressive manner for factors 1 to 5, to emulate different SNRs of the signal, despite the fact that this is not a common practice in telecom applications. A factor of 5 is used in the results, except if stated otherwise. The digital samples are sent offline to a off-line digital FFE with 55 taps. For the classification, 10000 symbols are used for training and 45000 for testing.

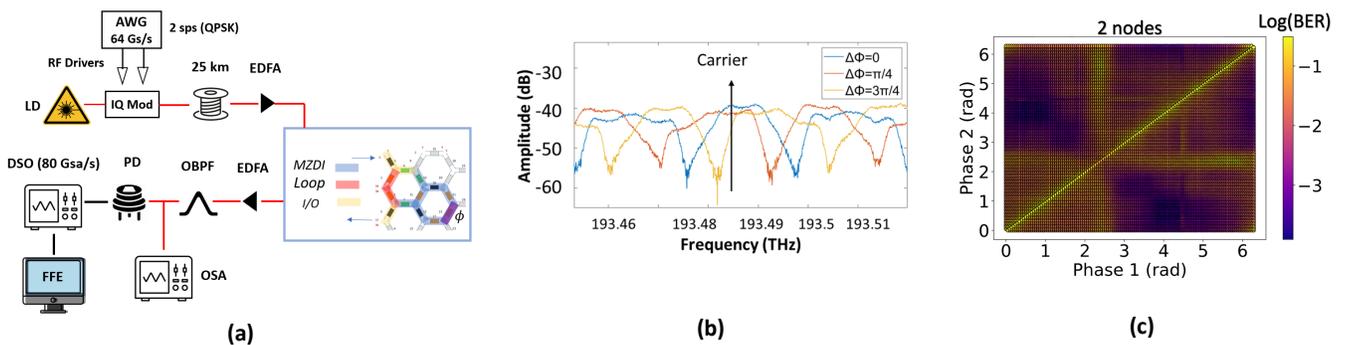

Fig. 2: (a) The experimental setup for the HA-PUF. The inset presents the programmed ROSS node on the SmartLight. (b) Different transfer functions for different phase configurations. (c) Multiple sub-FEC configurations of a ROSS structure with two nodes.

By changing $\phi$ at each MZDI, the spectrum of the optical filter is modified, thus providing different spectral components at its output, as is shown in Fig. 2.b. A total of 125 different phase configurations are used, resulting in an equal number of spectral responses. Each response is recorded repeatedly 60 times to evaluate the robustness of the system with respect to noise. Since the self-coherent receiver requires more than two ROSS nodes acting in parallel, for amplitude and phase extraction via the phase-to-amplitude mechanism, the responses are grouped in sets of $N_f = 2,3,4$ nodes emulating recurrent filters slicing the input in parallel. For two or more filters, multiple phase configurations provide sub-FEC BER ($< 1.25 \times 10^{-2}$), while the best performance was $3 \times 10^{-4}$ [16] (Fig. 2.c). By increasing the number of ROSS nodes to 3, equalization performance improves topping at $10^{-5}$, while now there is a total of 157,642 sets of phase combinations (challenges) corresponding to digital weights (responses), offering sub-FEC BER performance. This perspective allows to examine the variation in the weights generated by node-sets that have a minimum phase difference $\delta\phi_{\min}$ among them (phase threshold). To generate the digital keys, the weights are randomly projected to a higher dimensional plane of 5000 values, using a random binary fuzzy extractor [17]. This process generates binary keys that follow a uniform distribution [11] with bit precisions between 1 and 5 bits, based on the Gray encoding. Finally, a random subset of 256 bits is extracted from each binary response to form the secret key. The parameters for the fuzzy extractor are fixed to render fabrication imperfections the only source of uncertainty.

To achieve unpredictability, the Hamming distance (HD) between responses from different challenges needs to be high. HD is the number of bit-flips between bit series. Such a distribution of HDs is known as inter-HD and when normalized it is ideally centered to 0.5 (pure randomness). Here, it is calculated with the boot-strap method. In each step a sample of 5000 challenge-pairs is used to approximate the distribution. This process is repeated 1000 times to acquire adequate statistical properties. The final probability density function (PDF) is fitted with a mixture of Gaussians. To check the effect of different phase biases, 7 different distributions are computed for passive offset-pairs with phase thresholds equal to 0, 0.2, ..., 1.2 rad. For robustness, the HD between responses from the same challenge needs to be low. Such a distribution of HDs is known as the intra-HD. For optimum performance it must not overlap with the inter-HD distribution. To test this property, 60 evaluations from 20 uniformly sampled challenges are considered. All possible HDs per challenge are in total 1770, thus yielding $20 \times$

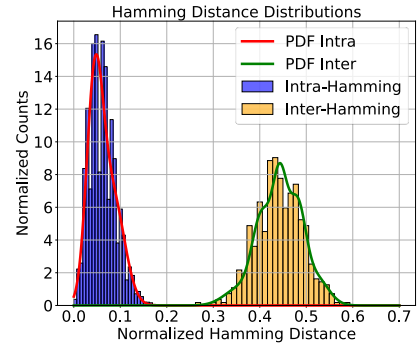

Fig. 3: The Intra-Hamming and Inter-Hamming distance distributions for 3 ROSS nodes and 3-bit precision.

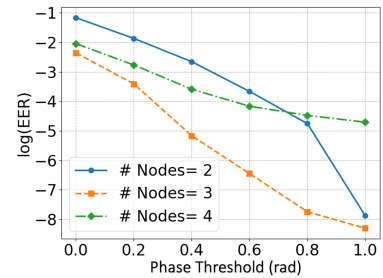

Fig. 4: The relation between the EER, the number of ROSS nodes and the phase threshold.

$1770 = 35400$ HDs. The PDF of intra-HD is again fitted with a mixture of Gaussians. Given the generated key, the authentication authority will accept/reject the remote device, according to the HD between the produced key and the expected key. Thus, a decision threshold $d_T$ is defined. Errors occur when a response from another device to the same challenge is close to the required key (false acceptance rate), as well as when the expected device produces a key highly different from the expected key due to noise effects (false rejection rate). False acceptance rate is the probability $P_{FA}(d_T) = P_{Inter}(d < d_T)$, whereas false rejection rate is the probability $P_{FR}(d_T) = 1 - P_{intra}(d < d_T)$. To balance unpredictability and robustness, a threshold $d_{EER}$ is established that equates $P_{FA}(d_{EER}) = P_{FR}(d_{EER})$, known as the equal error rate (EER) [6]. In this case, the two probabilities must be lower than $10^{-6}$ for appropriate performance and lower than $10^{-9}$ for excellent performance [6]. The EER is approximated by the fitted PDFs for the inter-HD and intra-HD distributions. For 3-bit precision and a phase threshold equal to 0.6 rad, the two distributions are depicted in Fig. 3.

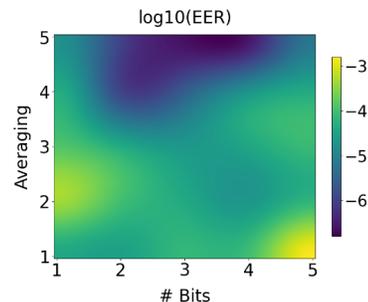

Fig. 5: Relation between averaging, the bit precision and the EER, for 3 nodes and 0.6 rad phase threshold.

The fitting PDF for both distributions is a mixture of two Gaussians, where the fitting parameters are the mean value and the standard deviation per Gaussian. They are clearly separated, with an EER equal to $2 \times 10^{-7}$. The relation between the number of nodes, the phase threshold and the EER is depicted in Fig. 4, for 3-bit quantization. Low phase threshold values correspond to low effective phase differences ($\delta\phi_{eff}$). As the phase threshold increases, the EER rapidly decreases, reaching values lower than $10^{-6}$ for 3 nodes. In the case of 2 nodes, the simplicity of the architecture limits the entropy of the PUF. On the contrary, in the case of 4 nodes, more experimental noise is introduced to the FFE due to the additional node, thus degrading the robustness of the PUF (broadening of the intra-HD distribution). A similar relationship between the number of nodes and the SNR has been observed in [10]. The relation between the averaging, the number of bits and the EER is presented in Fig. 5, for the case of 3 ROSS nodes and a phase threshold of 0.6 rad. As the averaging increases, the performance improves due to the improvement of the SNR. With respect to the bit precision, low values provide coarse quantization that increases robustness but also decreases the difference between weight values provided by different challenges. On the other hand, precision higher than 4 bits increases the effect of noise in the form of bit-flips, since the entropy of the PUF mechanism is lower than 5 bits. Thus, optimum performance is achieved in the range between 2-4 bits.

In conclusion, the provided results validate the PHA-PUF concept, where the unpredictable passive offsets are mapped to the digital weights of the post-processing module, when challenged with a set of optimized phases that achieve dispersion compensation. The PHA under study is a ROSS architecture implemented on a commercial reconfigurable integrated silicon photonic chip. In this context, parameters such as the complexity of the architecture - number of physical nodes-, the quantization level and the SNR are examined. Results can be further improved, by leveraging more degrees of freedom (phase configurations, such as the phase of the loop) or by using, advanced fuzzy extractors that include error correction modules [6]. The presented concept although realized for the ROSS architecture, is also compatible with other PHAs, such as convolutional layers [4], feedforward layers [18] and other RC topologies [19]. Therefore, this technology paves a way to combine high-speed photonic processing with secure authentication in the upcoming era of trusted ICT systems in the Internet of Everything.

**Funding.** This work was funded in part by PROMETHEUS (ID: 101070195), QPIC1550 (ID: 101135785) Horizon Europe projects. G. Sarantoglou is supported by the Project QUASAR which is implemented in the framework of H.F.R.I call "Basic research Financing (Horizontal support of all Sciences)" under the National Recovery and Resilience Plan "Greece 2.0" funded by the EU – Next Generation EU No: 016594. F. Da Ross is supported by Villum Foundations (YIP project OPTIC-AI no. VIL29344.

**Disclosures.** The authors declare no conflicts of interest.

**Data Availability.** Data may be obtained from the authors upon reasonable request